\documentclass[physrev, twocolumn, amsmath, amssymb, floats, floatfix,longbibliography]{revtex4-2}

\DeclareMathOperator{\sign}{sign}
\DeclareUnicodeCharacter{2212}{-}

\usepackage{graphicx}
\usepackage{dcolumn}
\usepackage{bm}
\usepackage{hyperref}
\hypersetup{
    colorlinks,
    citecolor=black,
    filecolor=black,
    linkcolor=black,
    urlcolor=black}
\usepackage{braket}
\usepackage{soul}
\usepackage{enumerate}

\begin{document}

\title{Topological classification of one-dimensional chiral symmetric interfaces}
\author{Harry MullineauxSanders}
\affiliation{SUPA, School of Physics and Astronomy, University of St.~Andrews, North Haugh, St.~Andrews KY16 9SS, United Kingdom}
\author{Bernd Braunecker}
\affiliation{SUPA, School of Physics and Astronomy, University of St.~Andrews, North Haugh, St.~Andrews KY16 9SS, United Kingdom}

\date{\today}

\begin{abstract}

We address the topological classification of one-dimensional chiral symmetric interfaces embedded into a two-dimensional substrate. A proof of the validity of a topological classification based on
the Green’s function by explicit evaluation of the topological invariant is presented. Further, we show that due to entanglement between the in-gap modes and the substrate, the full physics of the substrate that is contained in the Green's function is required. This is done by considering a classification scheme derived from the reduced ground state projector, for which we show that an uncritical handling produces erroneous changes in the topological index due to entanglement driven gap closures. We illustrate our results by applying them to a tight-binding model of a spiral magnetic interface in a $s$-wave superconductor.
\end{abstract}
\maketitle
\textit{Introduction. }The realisation of topological quantum states has been a major aim of condensed matter physics for a number of years.
Most prominent is obtaining robust Majorana modes \cite{kitaev_unpaired_2001,read_paired_2000,sato_topological_2017}, as they would provide the first steps toward topological quantum computation \cite{kitaev_fault-tolerant_2003,pachos_introduction_2012}.
In the absence of intrinsic topological properties in a material \cite{sharma_comprehensive_2022}, artificially created topological structures \cite{oreg_helical_2010,nadj-perge_proposal_2013} have emerged as promising candidates for tunable topological states, and specifically chains of magnetic impurities on standard s-wave superconductors, have been of great interest recently.

\begin{figure}
    \centering
    \includegraphics[width=0.45\textwidth]{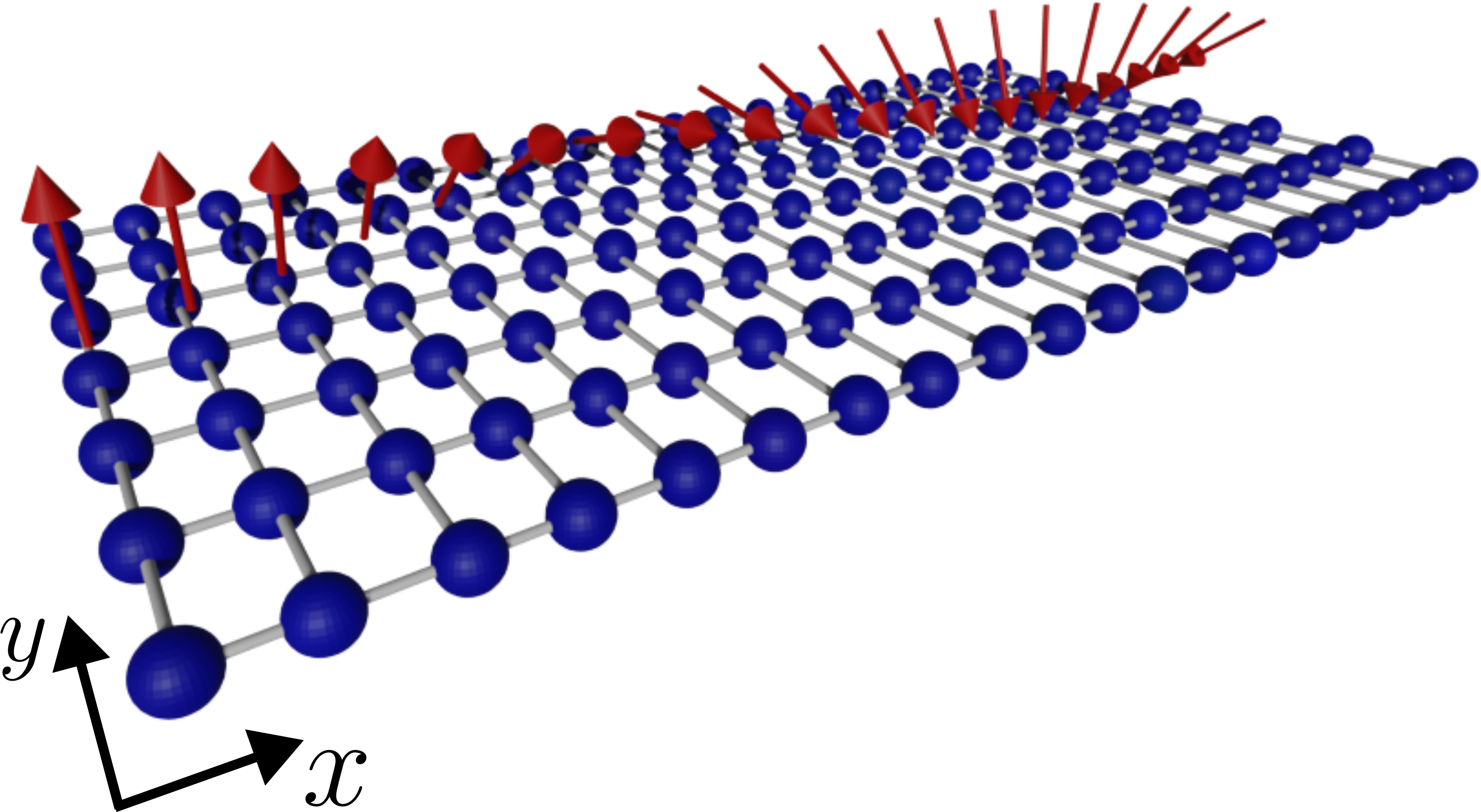}
    \caption{Sketch of a spiral magnetic interface model: A square lattice tight-binding model of a $s$-wave superconducting substrate with a chain of spiralling magnetic impurities along the $x$ direction.}
    \label{System-schematic}
\end{figure}

Magnetic impurities locally break Cooper pairs \cite{reis_self-organized_2014,meng_superconducting_2015} and create Yu-Shiba-Rusinov bound states \cite{yu_luh_bound_1965,shiba_classical_1968,rusinov_superconductivity_1969}. Chains of impurities, such as shown in Fig.~\ref{System-schematic}, create in-gap bands \cite{shiba_classical_1968} which can be tuned to a topological phase through scattering strength and magnetic alignment. The topological properties of such magnetic interfaces have been determined by diagonalisation of tight-binding models \cite{nadj-perge_proposal_2013,reis_self-organized_2014, christensen_spiral_2016,li_topological_2014}, by first principles calculation of Green's functions \cite{nyari_topological_2023,laszloffy_topological_2023,sticlet_topological_2019}, by analytic determination of Green's functions with a Debye frequency cut-off \cite{mier_calculations_2021}, by effective tight-binding hybridisation of impurity states \cite{choy_majorana_2011,pientka_topological_2013,pientka_unconventional_2014,weststrom_topological_2015,brydon_topological_2015,heimes_majorana_2014,poyhonen_topological_2016}, and by modelling one-dimensional (1D) magnetic wires with proximity induced superconductivity \cite{andolina_topological_2017,peng_strong_2015}. For densely packed chains, the effect of the reduced dimensionally of the embedded interface on band structure and topological properties has been investigated \cite{wang_impurity_2004,carroll_subgap_2021,carroll_subgap_2021-1,sedlmayr_analytical_2021,Sedlmayr_2022}. There has also been significant experimental effort in fabricating and studying these interfaces, some of which have shown signs of topological physics, however this evidence remains inconclusive \cite{nadj-perge_observation_2014,ruby_end_2015,menard_coherent_2015,pawlak_probing_2016,feldman_high-resolution_2017,ruby_exploring_2017,schneider_precursors_2022,liebhaber_quantum_2022,kim_toward_2018,schneider_controlling_2020}.

As the topological properties are created at the interface, a characterisation through effective 1D Hamiltonians obtained by tracing out the transverse spatial dimensions would be most convenient. However, this is only valid for in-gap modes tightly confined to the interface \cite{carroll_subgap_2021-1}, and typically not realised in experiment \cite{menard_coherent_2015}.
Otherwise the bound modes have sufficient entanglement between the degrees of freedom in the unit cell and in space, such that essential information for topological classification is lost through spatial traces \cite{carroll_subgap_2021-1}. A usable 1D Hamiltonian must thus rely on a quantity that does not eliminate essential spatial information, and it has been proposed that this can be achieved through local Green's functions \cite{carroll_subgap_2021-1,sedlmayr_analytical_2021} yet only with circumstantial evidence.

In this paper, we prove that this approach is generally valid for any 1D interface embedded into a trivial substrate with a chiral symmetry under the assumption that the real space Green's function of the substrate contains no zeros. Thus we provide a fast, versatile and numerically and analytically tractable classification scheme. As an illustration we determine the topological phase diagram of a tight-binding model of a 1D magnetic interface in a two-dimensional (2D) superconductor. We demonstrate further that this method is resilient against the spatial entanglement of the bound states by comparing with the reduced ground state projector approach \cite{tuegel_embedded_2019}, and that the latter requires careful handling to avoid artefacts in the parameter regions of most interest.

\textit{Topological Hamiltonian.} We consider a gapped translationally invariant 2D substrate with Hamiltonian $\mathcal{H}_0(k_x,k_y)$, for momenta $k_x,k_y$, with the gap centred about energy $\omega=0$. A scattering interface is introduced as a strip along the $x$ direction, bounded in the $y$ direction. For simplicity, we assume the interface to be localised at $y=0$, however generalisation to wider interfaces is immediate by considering enhanced unit cells along $y$ of the width of the interface. Assuming translational invariance along $x$, the scattering Hamiltonian is $\mathcal{H}_V(k_x)=H_V(k_x)\otimes P_{y=0}$, where $P_{y=0}$ projects onto the interface and $H_V$ acts on the remaining degrees of freedom.

We denote by $g(\omega,k_x,y)$ the retarded Green's function at energy $\omega$ of $\mathcal{H}_0(k_x,k_y)$ after the partial Fourier transform $k_y\rightarrow y$.
Then the total Green's function of $\mathcal{H}=\mathcal{H}_0+\mathcal{H}_V$ with broken translational invariance in the $y$ direction is given by Dyson's equation \cite{economou_greens_2006}
\begin{multline}
    G(\omega,k_x,y,y')
	=
	\\
	g(\omega,k_x,y-y')
	+g(\omega,k_x,y)T(\omega,k_x)g(\omega,k_x,-y'),
    \label{Greens-function-dyson-series}
\end{multline}
with the $T$ matrix
\begin{gather}
    T(\omega,k_x)=[H_V^{-1}-g(\omega,k_x,0)]^{-1}.
    \label{T-matrix-def}
\end{gather}
The poles of $T(\omega,k_x)$ at fixed $k_x$ provide the bound state energies, and their residues information on their wavefunctions. Consequently, $T$, and thus $G$, are expected to contain all topological information, and it was proposed \cite{carroll_subgap_2021-1,sedlmayr_analytical_2021} that
\begin{gather} \label{eq:Htop}
    H^{top}(k_x,y)=-[G(\omega=0,k_x,y,y)]^{-1}
\end{gather}
forms at $y=0$, a suitable topological Hamiltonian for topological classification. Similar Hamiltonians have been used for interacting topological materials \cite{volovik_universe_2009,wang_simplified_2012,wang_topological_2013}.

We shall now prove that $H^{top}$ is generally applicable for the topological classification for systems with chiral symmetries.
This relies on the demonstration that $H^{top}$ possesses the same global symmetries as the full 2D Hamiltonian. Let $\mathcal{H}(k_x)$ be the full Hamiltonian at fixed $k_x$, considered as a matrix acting on all other degrees of freedom including $y$. As $\mathcal{H}(k_x)$ is gapped, $G(\omega,k_x)=[\omega \openone-\mathcal{H}(k_x)]^{-1}$ is nonsingular at $\omega=0$. If $\Tilde{\Lambda}$ is a symmetry operator such that $\Tilde{\Lambda}\mathcal{H}(k_x)\Tilde{\Lambda}=\pm \mathcal{H}(\pm k_x)$ for a given combination of $\pm$ signs, then $\Tilde{\Lambda}G(0,k_x)\Tilde{\Lambda}=\pm G(0,\pm k_x)$ holds as well \cite{gurarie_single-particle_2011}. If $\Tilde{\Lambda}$ is global such that $\Tilde{\Lambda}=\Lambda\otimes\openone_y$ where $\openone_y$ is the identity for the $y$ coordinate and $\Lambda$ acts on the remaining degrees of freedom \cite{tuegel_embedded_2019}, then it commutes with the projection onto $y$ and
\begin{equation} \label{eq:symm}
	\Lambda^{-1}G(\omega=0,k_x,y,y)\Lambda=\pm G(\omega=0,\pm k_x,y,y).
\end{equation}
Therefore, $H^{top}(k_x,y) = -[G(\omega=0,k_x,y,y)]^{-1}$ satisfies all the global symmetries of the full Hamiltonian.
Since furthermore the interface region is bounded in the $y$ direction the following theorem holds (see Appendix \ref{appendix-global-symmetry-proof}):
\begin{gather}
	\nonumber
    \text{$\mathcal{H}=\mathcal{H}_0+\mathcal{H}_V$ possesses a global symmetry $\Lambda\otimes\openone_y$}
	\\
    \label{symmetry-theorem}
	\Leftrightarrow
	\\
	\nonumber \text{$\Lambda$ is a symmetry of $g(0,k_x,y)$ and $H_V$ individually}.
\end{gather}
It follows from Eq.~(\ref{T-matrix-def}) that the $T$ matrix also possesses the same symmetry. This means in any symmetry class, we can define a topological invariant individually for the topological Hamiltonian, the substrate Green's function, the $T$ matrix and the scattering Hamiltonian. While the $\mathcal{H}(k_x)$ and $H^{top}(k_x,y)$ share the same symmetries, it is still possible for the topological Hamiltonian at a fine tuned $y$ to have an enhanced symmetry which may effect the symmetry protection of its invariant.

We now specialise to chiral 1D symmetry classes which are relevant to magnetic impurity-superconductor interfaces \cite{tewari_topological_2012,sedlmayr_analytical_2021,carroll_subgap_2021-1}. In these classes, $\mathcal{H}(k_x)$ has a unitary Hermitian symmetry $\mathcal{C}$ such that $\mathcal{C}\mathcal{H}(k_x)\mathcal{C}=-\mathcal{H}(k_x)$. The appropriate topological invariant is the winding number, $W$, calculated by writing the $H^{top}$ in the eigenbasis of $\mathcal{C}$ \cite{chiu_classification_2016},
\begin{gather}
\label{chiral-decomposition-def}
    H^{top}(k_x,y)=\begin{pmatrix}
        0&h(k_x,y)\\
        h^\dagger(k_x,y)&0
    \end{pmatrix},
\end{gather}
with $h(k_x,y)$ called the chiral decomposition, such that
\begin{gather}
    W(y)=\frac{1}{2\pi i}\int_{-\pi}^{\pi}dk_x\partial_{k_x}\log\{\det[h(k_x,y)]\}.
    \label{winding-number-definition}
\end{gather}
Theorem~(\ref{symmetry-theorem}) ensures that $g(\omega,k_x,y)$, $T(\omega=0,k_x)$, and $H_V$ individually admit simultaneous chiral decompositions as in Eq.~(\ref{chiral-decomposition-def}).
Since $H^{top}(k_x,y=0) = - H_V T^{-1}(0,k_x) [g(0,k_x,0)]^{-1}$, we obtain (see Appendix \ref{appendix-winding-number})
\begin{equation}
    W(y=0)=W_{g^{-1}(y=0)}-W_{T^{-1}}+W_{H_V},
    \label{topological-hamiltonian-winding-number-formula}
\end{equation}
which is the central result of this paper.
Here $W_A$ represents of the winding number obtained from Eq.~(\ref{winding-number-definition}) with $h(k_x,y)$ replaced by the chiral decomposition of $A=[g(\omega=0,k_x,y)]^{-1}$, $[T(\omega=0,k_x)]^{-1}$, and $H_V$ respectively.

We note that $W_{g^{-1}(y=0)}$ can change its value only if a substrate pole passes through $\omega=0$ or if the Green's function has a zero. The zeros do not reflect any gap closure of the full Hamiltonian and so create a fictitious phase boundary. However previous work has suggested that such zeros of the partially Fourier transformed, in-gap Green's function are related to the topology of the substrate and are not present in trivial phases \cite{slager_impurity-bound_2015,diop_impurity_2020}. We shall proceed under the assumption that the substrate Green's function does not contain a zero.

For a trivial substrate in the absence of zeros and trivial (including on-site) scattering, $W_{g^{-1}(y=0)}=W_{H_V}=0$, and therefore
\begin{equation}
    W(y=0)=-W_{T^{-1}}.
    \label{T-matrix-winding-number}
\end{equation}
The invariant only becomes undefined and changes value when $\det\{[T(\omega=0)]^{-1}\}=0$, matching when the in-gap bands touch.

For a general $H_V$ with $k_x$ dependence, a gap closure could occur. These gap closures cause zeros in the $T$ matrix and can change its winding number \cite{gurarie_single-particle_2011}. However, the relative negative sign between the two terms in Eq.~(\ref{topological-hamiltonian-winding-number-formula}) means any change caused by the appearance of such a zero is compensated by the change in winding number of the scattering Hamiltonian (see Appendix \ref{appendix-winding-number}).

Based on these results we prove in Appendix \ref{appendix-proof-equivalence} that, for a trivial substrate without zeros, the topological Hamiltonian is trivial if and only if the Hamiltonian is trivial and as a consequence the the winding number of the Hamiltonian and topological Hamiltonian must be equal up to a possible sign that is global to the phase diagram. We have therefore proven that for a trivial substrate, the topological Hamiltonian proposed in \cite{carroll_subgap_2021-1,sedlmayr_analytical_2021} provides the correct topological classification when $y$ is tuned to the interface.

The classification in the case of a topologically nontrivial substrate via the topological Hamiltonian remains outstanding.
$H^{top}$ contains a dimensional reduction from a higher $D$ dimensional substrate to a lower $d$ dimensional interface. However, the allowed values of topological invariants change between dimensions \cite{chiu_classification_2016} and symmetries allowing for a nontrivial system in $D$ dimensions may impose only trivial states in $d$ dimensions. Therefore $[g(\omega=0,k_x,y)]^{-1}$ does not necessarily form a good topological Hamiltonian of the substrate, whereas a trivial state remains trivial after any dimensional reduction as it remains always amenable to the atomic limit. In addition, as proposed in Refs.~\cite{slager_impurity-bound_2015,diop_impurity_2020}, the substrate Green's function should generically have zero eigenvalues at some in-gap frequency. This reflects that at arbitrarily large scalar potential, an edge is created which should have a bound mode due to the bulk boundary correspondence. Such zeros could change the invariant of the topological Hamiltonian without a gap closure and could give inaccurate results.

When $y\neq 0$, the winding number becomes (see Appendix \ref{appendix-winding-number})
\begin{gather}
	\label{eq:W(y)}
    W(y)=W_{g^{-1}(y)} + W_{g^{-1}(-y)} -W_{T^{-1}}\\\nonumber
    -\frac{1}{2\pi i}\int_{-\pi}^{\pi}dk
	\partial_k\log\Bigl\{\det\bigl[\openone+\Tilde{g}^{-1}(-y)\Tilde{T}^{-1\dagger}\Tilde{g}^{-1}(y)\Tilde{g}(0)\bigr]\Bigr\},
\end{gather}
where $\Tilde{g}^{-1}(y)$ and $\Tilde{T}^{-1}$ are the chiral decompositions of $[g(\omega=0,k_x,y)]^{-1}$ and $[T(\omega=0,k_x)]^{-1}$, respectively. The final term can be singular and possess nontrivial winding. Indeed, as $|y|\rightarrow\infty$ the zeros of $G(\omega=0,k_x,y,y)$ cancel the poles of $T(\omega=0,k_x)$ (see Appendix \ref{appendix-winding-number}) and the winding of the $T$ matrix is perfectly cancelled. Therefore $H^{top}(|y|\rightarrow\infty)$ is trivial, as observed already in Ref.~\cite{carroll_subgap_2021-1}. As such only $y=0$ should be considered for topological classification as $y\neq0$ or integrals over $y$ may produce erroneous results.

Equation~(\ref{topological-hamiltonian-winding-number-formula}) is a general result applicable to any gapped substrate
with any interface that has a chiral symmetry. Furthermore, it applies also to interfaces of larger width, by extending the matrix sizes to include the $y$ coordinates of the interface too.
The key appeal of this method is that it is calculated from analytically tractable objects, well defined in the system size $L\rightarrow\infty$ limit, removing any finite size effects. As only the local Green's function is required, this approach is applicable to first principals calculations \cite{nyari_topological_2023,laszloffy_topological_2023,sticlet_topological_2019,feldman_high-resolution_2017}. It also opens the door to connecting this classification to local response functions that are experimentally accessible.

\begin{figure}
    \centering
    \includegraphics[width=0.95\columnwidth]{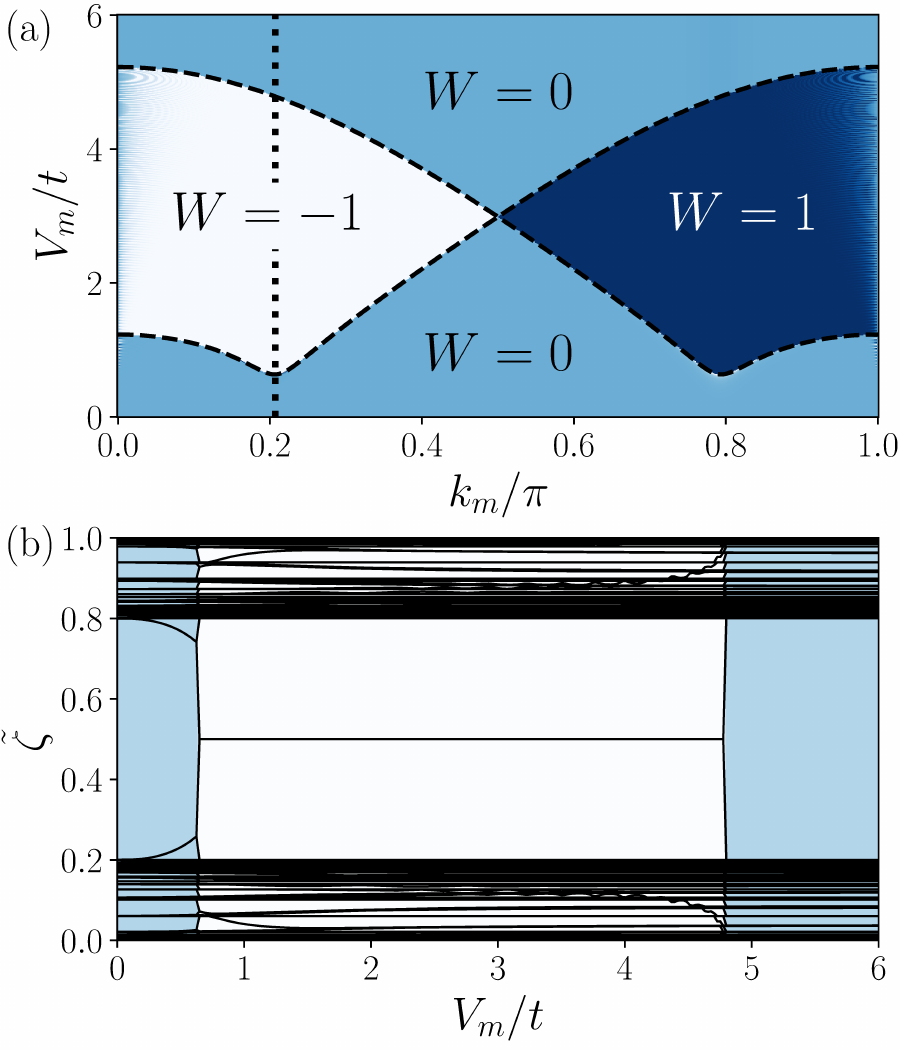}
    \caption{(a) Winding number of $H^{top}(k_x,y=0)$, with values represented by the background colour, for $\Delta=0.1t$ and $k_F=0.65$. Black dashed lines show the phase boundaries [Eq.~\eqref{phase-boundaries}]. At $k_m=0,\pi$ the model is gapless and changes symmetry class. The fuzzy pattern nearby arises from numerical instabilities caused by tiny gaps. The minima of the lower phase boundary are at $k_m=k_F,\pi-k_F$.
	(b) Entanglement spectrum $\Tilde{\zeta}$ at $k_m=k_F$ for $V_m$ varying along the dotted line in (a), computed for a $50 \times 101$ lattice with an entanglement cut made perpendicular to the impurity chain splitting the system in two. The background colour is as in (a). Edge modes, characterised by $\Tilde{\zeta}=1/2$, appear exactly when the winding number of the topological Hamiltonian is nontrivial.}
    \label{Fig:Top-Hamiltonian-Phase-Diagram}
\end{figure}

\textit{Application to magnetic interface in a superconductor.} We demonstrate the usefulness of this approach with the topological classification of a chain of spiralling classical magnetic impurities embedded into a tight-binding model of a BCS ($s$-wave) superconductor, such as shown in Fig.~\ref{System-schematic}.
Similar to Refs.~\cite{nadj-perge_proposal_2013,pientka_topological_2013,weststrom_topological_2015,carroll_subgap_2021,carroll_subgap_2021-1} we assume the impurities to form a scattering interface at $y=0$ with planar magnetisation $\textbf{M}(x)=V_m(\cos(2k_mx),\sin(2k_mx),0)$, where $V=V_m$ sets the scattering strength, and $k_m$ the period of the spiral which is considered as a free parameter and not determined self-consistently \cite{braunecker_interplay_2013,kim_helical_2014,klinovaja_topological_2013,vazifeh_self-organized_2013,reis_self-organized_2014,schecter_spin-lattice_2015,schecter_self-organized_2016}.
Translational symmetry along $x$ can be restored by a gauge transformation on the electron creation operator $c^\dagger_{x,y,\sigma}\rightarrow e^{ik_m\sigma x}c^\dagger_{x,y,\sigma}$, for spin $\sigma=\uparrow,\downarrow=+,-$ \cite{braunecker_spin-selective_2010}. This maps the magnetisation to $\mathbf{M}=V_m (1,0,0)$ but spin splits the bands by the shifts $k_x\rightarrow k_x+\sigma k_m$.
In $(k_x,y)$ space the components of the Hamiltonian $\mathcal{H}=\mathcal{H}_0+\mathcal{H}_V$ are then written as
\begin{align}
    &\mathcal{H}_0
	=
	\sum_{k_x,y,\sigma}
	\bigl[-2t\cos(k_x+\sigma k_m)-\mu\bigr] c^\dagger_{k_x,y,\sigma}c_{k_x,y,\sigma}
\label{Hamiltonian-fourier-transform}
\\ \nonumber
    &-t\sum_{k_x,y,\sigma}
	\Bigl[
		\bigl(c^\dagger_{k_x,y+1,\sigma}c_{k_x,y,\sigma}+\Delta c^\dagger_{k_x,y,\uparrow}c^\dagger_{-k_x,y,\downarrow} \bigr)
		+ \text{h.c.}
	\Bigr],
\end{align}
and $\mathcal{H}_V=V_m\sum_{k_x}c^{\dagger}_{k_x,0,\uparrow}c_{k_x,0,\downarrow}+\text{h.c.}$,
with hopping integral $t$, pairing $\Delta$, and chemical potential $\mu$. The coordinates $x,y$ label a square lattice with lattice constant $a=1$. We consider a constant $\Delta$ as this allows for analytic solution for the phase boundaries, and as its self-consistent adjustment near the impurities has little effect (see Appendix \ref{sec:self_consistent}). We finally note that the substrate Green's function cannot have a zero for any nonzero $\Delta$.

For $k_m\neq n\pi$ where $n\in\mathbb{Z}$, the model is in the BDI symmetry class \cite{chiu_classification_2016}, with chiral $\mathcal{C}=\tau^y\sigma^x$, particle-hole $\mathcal{P}=i\tau^xK$, time reversal $\mathcal{T}=\tau^z\sigma^xK$, and inversion symmetry $\mathcal{I}=\tau^z\sigma^x$ operators \cite{varjas_qsymm_2018}, for spin $\sigma^i$ and particle-hole $\tau^i$ Pauli matrices, and complex conjugation $K$.
The calculation of $g(\omega,k_x,y)$ follows Ref.~\cite{carroll_subgap_2021} and is performed in Appendix \ref{sec:Fourier}, from which we obtain $T(\omega,k_x)$. The poles of $T(0,k_x)$ provide the gap closures of the in-gap bands that mark the topological phase boundaries,
and occur at the high symmetry points $k_x=0,\pi$ for the scattering strengths [see Appendix \ref{sec:phase_diagram}]
\begin{equation}
    V^*_{m,k_x}=\pm\left\{\bigl[g^{\uparrow}_{11}(k_x)\bigr]^2+\bigl[g_{12}^{\uparrow}(k_x)\bigr]^2\right\}^{-1/2},
    \label{phase-boundaries}
\end{equation}
where $g^\uparrow_{11}$ and $g^\uparrow_{12}$ are the ($\uparrow$,particle)-($\uparrow$,particle) and ($\uparrow$,particle)-($\downarrow$,hole) terms of $g(\omega=0,k_x,y=0)$. In the continuum limit, $V^*_{m,k_x=0}$ reproduces the result of Ref.~\cite{carroll_subgap_2021-1}. Gap closures occur only at $k_x=0,\pi$, and so $W$ is restricted to $W=0,\pm1$. Using $H^{top}$ and Eq.~\eqref{T-matrix-winding-number} the parity of $W$ can be written as \cite{tewari_topological_2012}
\begin{equation} \label{eq:parity}
    (-1)^{W(y=0)}=\prod_{k_x=0,\pi}\sign\left[(V^*_{m,k_x})^{-2}-(V_m)^{-2}\right],
\end{equation}
and the phase is non-trivial for $\min( V_{m,k_x=0}^*,V_{m,k_x=\pi}^*)<V_m<\max( V_{m,k_x=0}^*,V_{m,k_x=\pi}^*)$. If instead $k_m=n\pi$, the spin sectors decouple and $\mathcal{H}$ block diagonalises into two $s$-wave Hamiltonians with each block subject to a local Zeeman shift $V_m\sigma^z$. Each block is in the trivial CI class such that always $W=0$ \cite{chiu_classification_2016}.

The resulting phase diagram is shown in Fig.~\ref{Fig:Top-Hamiltonian-Phase-Diagram}(a). To further corroborate our results, in Fig.~\ref{Fig:Top-Hamiltonian-Phase-Diagram}(b) we show the entanglement spectrum $\Tilde{\zeta}$ as $V_m$ is varied along the vertical dashed line in Fig.~\ref{Fig:Top-Hamiltonian-Phase-Diagram}(a) obtained from an entanglement cut made perpendicular to the chain \cite{hughes_inversion-symmetric_2011,peschel_calculation_2003,alexandradinata_trace_2011}. The entanglement spectrum is defined as the spectrum of the ground state projector $P_{GS}=\sum_{E_N<0}\ket{E_N}\bra{E_N}$ for eigenstates of the Hamiltonian $\ket{E_N}$ whose gap quantifies the degree of entanglement between two halves of the system. An edge is introduced into the system by restricting the position index to one half of the system. In a trivial phase $\Tilde{\zeta}$ is gapped about $1/2$, whereas in a nontrivial phase there is a doubly degenerate value $\Tilde{\zeta} = 1/2$ \cite{hughes_inversion-symmetric_2011}. This matches exactly our phase diagram.

\begin{figure*}
    \centering
    \includegraphics[width=0.95\textwidth]{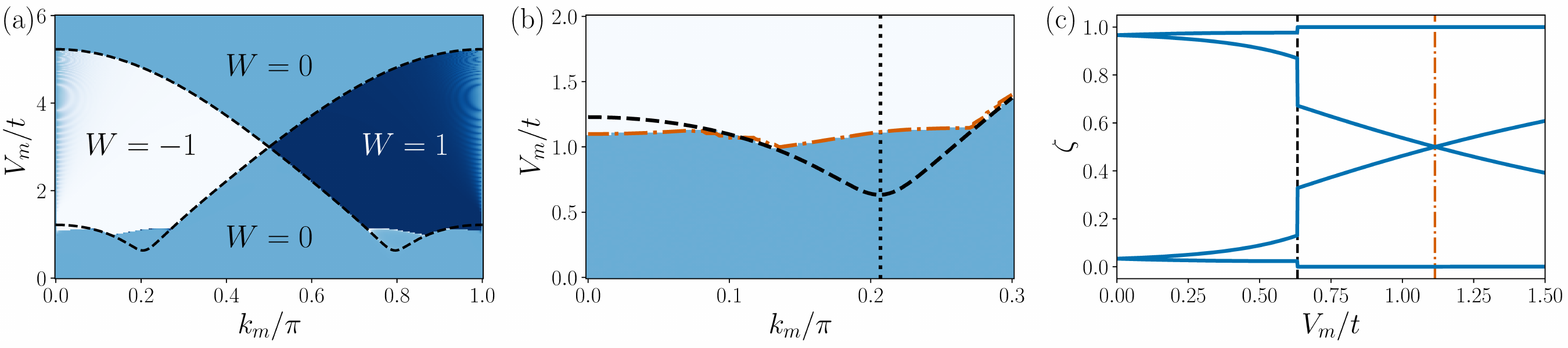}
    \caption{(a) Winding number of $1-2\Pi^{\mathcal{R}}$, the projector restricted to the impurity site at $y=0$, for $N_y=101$, $k_F=0.65$, and $\Delta=0.1t$. The black dashed lines show the phase boundaries [Eq.~\eqref{phase-boundaries}]. The classification gives the incorrect topological invariant, overshooting the phase boundary for $k_m\approx0$ and undershooting it for $k_m\approx k_F$. The fuzzy pattern $k_ma=0,\pi$ arises from numerical instability near gap closures.
	(b) Same as (a) zoomed on $k_m\approx k_F$. The orange dash-dotted line shows where the gap of the projector closes, matching exactly where the winding number changes.
	(c) Entanglement spectrum $\zeta$ of the projector $P^\mathcal{R}$ reduced to just $y=0$, at $k_x=0$, $k_m=k_F$ as $V_m$ is varied along the dotted line in (b).
	The black dashed and orange dash-dotted lines mark the boundaries as in (b). The discontinuity in $\zeta$ (black line) mark the true phase transition, whereas the erroneous phase transition by the projector appears where the $\zeta$ gap closes instead (orange line).}
    \label{Fig:projector-classification}
\end{figure*}

\textit{Classification of magnetic interface model via reduced projector.} The topological Hamiltonian is a local probe, but through $G$ contains information on the entire band structure, not just on the in-gap states. It may thus be natural to ask how much involvement of the extended substrate states is required, and whether the topological properties are reproducible from the wavefunctions truncated to a small wedge around the interface. To provide an answer, we compare with the classification scheme proposed in Ref.~\cite{tuegel_embedded_2019} under spatial truncation for a $D$ dimensional trivial substrate with embedded $d<D$ dimensional topological model. If $\ket{GS}$ is the ground state, the classification is based on the projector \cite{peschel_calculation_2003}
\begin{gather}
    P_{i,j}^{\alpha,\beta}(k)=\bra{GS}c^\dagger_{k,i,\alpha}c_{k,j,\beta}\ket{GS},
    \label{GS-projector-def}
\end{gather}
where $k$ is the $d$ dimensional momentum in the translationally invariant directions, $i,j$ label the further spatial coordinates, and $\alpha,\beta$ the unit cell degrees of freedom. Spatial degrees of freedom can be traced out by restricting the projector to a spatial region $\mathcal{R}$ that contains the embedded system. This leaves a reduced projector $P^{\mathcal{R}}(k)$ \cite{alexandradinata_trace_2011} with eigenvalues $\zeta_i(k) \in [0,1]$ and eigenvectors $\ket{\zeta_i(k)}$.
The value $\zeta_i$ weights how much the eigenstate is supported in $\mathcal{R}$.
The crossing through $\zeta=1/2$ thus indicates a qualitative change of the entanglement between $\mathcal{R}$ and its complement, and the possible topological implication can be captured through the entanglement projector \cite{tuegel_embedded_2019}
$\Pi^{\mathcal{R}}(k)=\sum_{\zeta_i(k)>1/2}\ket{\zeta_i(k)}\bra{\zeta_i(k)}$,
which is in the same symmetry class as the Hamiltonian \cite{tuegel_embedded_2019}, such that $1-2\Pi^{\mathcal{R}}(k)$ acts like a spectrally flattened band Hamiltonian with generally a gap about zero and allows for the standard extraction of topological invariants \cite{chiu_classification_2016}.

However changes of the reduced projector's invariant do not necessarily match topological phase transitions. Such changes can also be caused by gap closures due to entanglement between $\mathcal{R}$ and its complement from sufficiently extended in-gap states, and not only by gap closures of $\mathcal{H}$.
Indeed, in Fig.~\ref{Fig:projector-classification}(a) we show the winding number of $1-2\Pi^{\mathcal{R}}$ for the magnetic interface model \cite{fidkowski_entanglement_2010}, with $\mathcal{R}$ reduced to just one site at $y=0$. The result does not match Fig.~\ref{Fig:Top-Hamiltonian-Phase-Diagram}(a), overshooting the phase boundary for $k_m\approx 0$ and undershooting for $k_m\approx k_F$, as shown in the closeup of Fig.~\ref{Fig:projector-classification}(b).
This error is reproduced for larger lattice sizes and in the entanglement spectrum, and only by enlarging $\mathcal{R}$ to a width of $\approx 6$ sites the correct phase diagram is recovered. The error is particularly significant as $k_m = k_F$ is of much interest for self-sustained magnetic spirals \cite{braunecker_interplay_2013,klinovaja_topological_2013,kim_helical_2014,vazifeh_self-organized_2013,reis_self-organized_2014}.

The projector's invariant can change if the projector is discontinuous or its gap closes. The former occurs when the Hamiltonian's gap closes and the ground state reconfigures. The latter is associated with $\zeta=1/2$ and represents maximal entanglement through maximisation of the entropy $S=-\zeta\log(\zeta)$ \cite{ryu_entanglement_2006}. Such modes occur when the in-gap wavefunctions spread across $\mathcal{R}$ and have strong correlation between $y$ and spin-Nambu degrees of freedom \cite{carroll_subgap_2021-1}. Figure~\ref{Fig:projector-classification}(c) shows they are indeed responsible for the erroneous phase diagram. That lower dimensional subsystems of gapped systems can become gapless has previously been reported in Dirac systems \cite{moghaddam_boiling_2022}.

The ground state projector classification therefore requires careful checking of the extent of the localised wavefunctions. Alternatively, every phase transition predicted by this method would need to be checked as to whether it is due to gap closing in the Hamiltonian or in the reduced projector. Both could become computationally costly. This underlines the significance of the role of the substrate in the dimensional reduction that is captured by a Green's function based method as put forward in this paper.

\textit{Conclusions.} We have shown that the topological classification of 1D interfaces embedded into a topologically trivial 2D substrate with chiral symmetry can be performed exactly by the local Green's function tuned to the interface. With Eq.~\eqref{topological-hamiltonian-winding-number-formula} we  provide an explicit formula disentangling the local scattering effects from the substrate, while maintaining the full spatial dependence of the substrate through the $T$ matrix. The topological Hamiltonian provides a very fast, analytically and numerically tractable classification, free of finite size effects or scaling with the size of the substrate. While it is necessary to assume that the substrate Green's function contains no zeros, previous work suggests this is generically true for a trivial substrate. The same studies suggest that zeros are generically present for topological substrates, complicating the classification of interfaces embedded into nontrivial substrates. We demonstrate further that the full spatial dependence is crucial for the classification, and care must be taken when using classification schemes relying on spatial truncation.
The proof of the applicability of a local Green's function classification to non-chiral symmetry classes and higher dimensions requiring Pfaffian or Chern invariants remains outstanding, but we believe such extensions are possible. A proof for Pfaffian invariants in 1D would cover class D superconductors and complete the classification of all non-trivial 1D symmetry classes \cite{budich_equivalent_2013,chiu_classification_2016}.

\textit{Acknowledgements.} The authors thank Joe Winter and Konner McNeillie for helpful discussions. HMS acknowledges studentship funding from EPSRC under Grant No. EP/W524505/1.
The work presented in this paper is theoretical. No data were produced, and supporting research data are not required.


\appendix


\section{\label{appendix-global-symmetry-proof}Proof of Theorem~\eqref{symmetry-theorem}}

We consider the 2D substrate with the embedded scattering interface, described by the Hamiltonian $\mathcal{H}(k_x) = \mathcal{H}_0(k_x) + \mathcal{H}_V(k_x)$. The scattering interface is assumed to be concentrated to site $y=0$, which could be a chain of impurities. Alternatively, as explained in the main text, if the interface has a larger width we can partition the $y$ coordinates into supercells of the width of the interface, such that integer $y$ label the position of the supercells, and the $y$ dependence within each supercell is taken into account as internal matrix index such as spin or particle-hole degrees of freedom. For an interface in the form of a chain, the supercell matches the unit cell. Since $\mathcal{H}_V(k_x)$ is restricted to $y=0$ we can write $\mathcal{H}_V(k_x)=H_V(k_x)\otimes P_{y=0}$, where $P_y$ projects onto supercell $y$.

Let $\{\ket{y}\}$ form an orthonormal basis of 1D wavefunctions localised at cell $y$ such that $\bra{y} \mathcal{A}(k_x) \ket{y'}$ reduces any operator $\mathcal{A}(k_x)$ to a matrix of internal degrees of freedom only.
We restrict ourselves to substrates that are fully translationally invariant, so $\bra{y}\mathcal{H}_0(k_x)\ket{y'}=\bra{y+n}\mathcal{H}_0(k_x)\ket{y'+n}$ for all $y,y',n$.
Consider a global symmetry of the Hamiltonian $\Tilde{\Lambda}$ such that $\Tilde{\Lambda}^{-1}\mathcal{H}(k_x)\Tilde{\Lambda}=\pm \mathcal{H}(\pm k_x)$, where the $\pm$ are chosen depending on the type of symmetry. As $\Tilde{\Lambda}$ is global, it acts site-wise \cite{tuegel_embedded_2019}, i.e. the symmetry is of the form $\Tilde{\Lambda}=\Lambda\otimes\openone_y$, where $\Lambda$ is an operator that acts on the degrees of freedom within a supercell and $\openone_y$ is the identity acting on the $y$ cell coordinate.
The action of the symmetry on the Hamiltonian gives
\begin{gather}
     \Tilde{\Lambda}^{-1}\bigl[\mathcal{H}_0(k_x)+\mathcal{H}_V(k_x)\bigr]\Tilde{\Lambda}
	 =
	 \pm\bigl[\mathcal{H}_0(\pm k_x)+\mathcal{H}_V(\pm k_x)\bigr].
     \label{symmetry-acting-on-Hamitonian}
\end{gather}
As $\mathcal{H}_V$ is local to the interface, there exists a $y^*>0$ such that $\bra{y'}\mathcal{H}_V(k_x)\ket{y''}=0$ for $|y'|,|y''|>y^*$, but also that $\bra{y'} \tilde{\Lambda}^{-1} \mathcal{H}_V(k_x) \tilde{\Lambda}\ket{y''}=0$ as $\tilde{\Lambda}$ acts only locally. Therefore for such $y',y''$
\begin{align}
    &\bra{y'}\Tilde{\Lambda}^{-1}\bigl[\mathcal{H}_0(k_x)+\mathcal{H}_V(k_x)\bigr]\Tilde{\Lambda}\ket{y''}
\nonumber\\
    &=\pm\bra{y'}\bigl[\mathcal{H}_0(\pm k_x)+\mathcal{H}_V(\pm k_x)\bigr]\ket{y''},
\end{align}
reduces to
\begin{align}
    \bra{y'}\Tilde{\Lambda}^{-1}\mathcal{H}_0(k_x)\Tilde{\Lambda}\ket{y''}
    =\pm\bra{y'}\mathcal{H}_0(\pm k)\ket{y''}.
\end{align}
Due to the translational invariance of $\mathcal{H}_0(k_x)$ the latter equality holds for any $y',y''$, and it follows that
\begin{align}
    \Tilde{\Lambda}^{-1}\mathcal{H}_0(k_x)\Tilde{\Lambda}
    =\pm\mathcal{H}_0(\pm k).
\end{align}
This means that $\tilde{\Lambda}$ is a symmetry for $\mathcal{H}_0(k_x)$ separately too. From Eq.~\eqref{symmetry-acting-on-Hamitonian} it follows in turn that
\begin{align}
    \Tilde{\Lambda}^{-1}\mathcal{H}_V(k_x)\Tilde{\Lambda}
    =\pm\mathcal{H}_V(\pm k),
\end{align}
so that $\tilde{\Lambda}$ is a symmetry also for $\mathcal{H}_V(k_x)$. It follows further that $\Lambda$ is a symmetry of $H_V(k_x)$.

Finally from Eq.~\eqref{eq:symm} in the main text, we obtain that if $\Tilde{\Lambda}$ is a symmetry of $\mathcal{H}_0(k_x)$, then $\Lambda$ is a symmetry of $g(\omega=0,k_x,y)$.

To prove the converse let us assume that $\Lambda$ is a symmetry of $g(\omega=0,k_x,y)$. Then $\Tilde{\Lambda}=\Lambda\otimes\openone_y$ is a symmetry of $\mathcal{H}_0(k_x)$ for the translationally invariant substrate. If furthermore $\Lambda$ is a symmetry of $H_V(k_x)$, then $\Tilde{\Lambda}$ is a symmetry of $\mathcal{H}_V(k_x)=H_V(k_x)\otimes P_{y=0}$. Therefore from Eq.~(\ref{symmetry-acting-on-Hamitonian}), $\Tilde{\Lambda}$ is a symmetry of $\mathcal{H}(k_x)$. This concludes the proof of Theorem~\eqref{symmetry-theorem}.

\section{\label{appendix-winding-number}Winding Number of Topological Hamiltonian}

The winding number can evaluated by writing the Dyson equation for Green's function [Eq.~\eqref{Greens-function-dyson-series}] as follows
\begin{gather}
    G(0,k_x,y,y')=g(0,k_x,y)T(0,k_x)g(0,k_x,-y')
\nonumber\\
	\times
	\Bigl\{ \openone +
	\bigl[g(0,k_x,-y')\bigr]^{-1} \bigl[T(0,k_x)\bigr]^{-1}
\nonumber\\
	\times
	\bigl[g(0,k_x,y)\bigr]^{-1} g(0,k_x,y-y')\Bigr\}.
\label{factorised-greens-function}
\end{gather}
where $\openone$ is the identity operator, the $y$ label supercells as explained in Sec.~\ref{appendix-global-symmetry-proof}, and the written $G, g$ and $T$ are matrices in all internal degrees of freedom of a supercell. The inversion of the substrate Green's function and the $T$ matrix is possible due to the assumption that both are gapped apart from at phase boundaries. The topological Hamiltonian [Eq.~\eqref{eq:Htop} in the main text] can then be written as
\begin{gather}
	H^{top}(k_x,y)=
	-
	\Bigl\{ \openone+ \bigl[g(0,k_x,-y)\bigr]^{-1} \bigl[T(0,k_x)\bigr]^{-1}
\nonumber\\
	\times \bigl[g(0,k_x,y)\bigr]^{-1}g(0,k_x,0)\Bigr\}^{-1}
\nonumber\\
	\times
	\bigl[g(0,k_x,-y)\bigr]^{-1} \bigl[T(0,k_x)\bigr]^{-1} \bigl[g(0,k_x,y)\bigr]^{-1}.
\end{gather}
To simplify the notation we shall omit the index $k_x$ in most of the following expressions.

\subsection{Case $y=0$}

Let us consider $y=0$ and set $g = g(0,k_x,0)$ and $T = T(0,k_x)$. The topological Hamiltonian then becomes
\begin{align}
    H^{top}(k_x)
	&=
	-
	\bigl(\openone+g^{-1}T^{-1}\bigr)^{-1} g^{-1} T^{-1} g^{-1}
\nonumber\\
    &=
	-
	\bigl[ \openone + g^{-1} \bigl( H_V^{-1}-g\bigr) \bigr]^{-1} g^{-1} T^{-1} g^{-1}
\nonumber\\
	&=
	- H_VT^{-1}g^{-1}.
\end{align}
Theorem~\eqref{symmetry-theorem} guarantees that $T$, $g$ and $H_V$, and therefore $H^{top}$, are all chiral symmetric, admit a simultaneous chiral decomposition, and can be written in the chiral eigenbasis
\begin{gather}
	H_V=\begin{pmatrix}
			0&h_V\\
			h_V^\dagger&0
    \end{pmatrix},\quad
	g^{-1}=\begin{pmatrix}
        0&\Tilde{g}^{-1}\\
        \Tilde{g}^{-1\dagger}&0
    \end{pmatrix},
\nonumber\\
    T^{-1}=\begin{pmatrix}
        0&\Tilde{T}^{-1}\\
        \Tilde{T}^{-1\dagger}&0
    \end{pmatrix},\quad
	H^{top}=\begin{pmatrix}
			0&h^{top}\\
			h^{top \dagger}&0
    \end{pmatrix},\quad
\end{gather}
where
\begin{gather}
    h^{top}=-h_V\Tilde{T}^{-1\dagger}\Tilde{g}^{-1}.
\end{gather}
The minus sign does not affect the winding number which can then be evaluated as
\begin{gather}
    W(y=0)
	=
	\frac{1}{2\pi i}\int_{-\pi}^{\pi }dk_x\partial_{k_x}\log[\det(h^{top})]
\nonumber\\
	=\frac{1}{2\pi i}\int_{-\pi}^{\pi}dk_x\partial_{k_x}
	\Bigl\{\log[\det(h_V)]+\log[\det(\Tilde{T}^{-1\dagger})]
\nonumber\\
	+\log[\det(\Tilde{g}^{-1})]\Bigr\}.
\end{gather}
Consider a general chiral symmetric operator $A$ with chiral decomposition $a$ and take a change of variable $z=\det(a)$ such that the $k_x$ integration over the first Brillouin zone becomes the $z$ integration over the closed contour
$\mathcal{Z} = \{ \, z = \det(a(k_x))\, |\, k_x\in[-\pi,\pi)\, \}$. We can then write the corresponding winding number in the form
\begin{gather}
    W_A=\int_{-\pi}^{\pi}\frac{dk_x}{2\pi i}\partial_{k_x}\log\bigl\{\det\bigl[a(k_x)\bigr]\bigr\}
	=\oint_\mathcal{Z}\frac{dz}{2\pi i z}.
\end{gather}
From this expression we see that if we take $a\rightarrow a^\dagger$, $\det(a)\rightarrow\det(a)^*$ then $\mathcal{Z}$ is inverted about the real axis and the contour is traversed in the opposite direction. This gives an overall minus to the winding number. Therefore the winding number of the topological Hamiltonian becomes
\begin{equation}
    W(y=0)=W_{g^{-1}}-W_{T^{-1}}+W_{H_V}.
    \label{winding-number-identity}
\end{equation}
A locally acting $H_V$ has a zero winding number. But general $k_x$ dependent $H_V(k_x)$ may lead to $W_{H_V} \neq 0$, and thus seemingly to a topological phase transition without gap closure. However, in such a case the change in $W_{H_V}$ is compensated by a simultaneous change in $W_{T^{-1}}$, so that a gap closure still remains mandatory. To see this, we should recall that $-W_{T^{-1}}+W_{H_V}$ represents the winding of $H_V T^{-1} = \openone - H_V g$, thus of $\det(\openone - h_V \tilde{g}^\dagger)$.
This means that the direct winding from $H_V$ is undone but the zeros of $T^{-1}$ are preserved. In case of doubt indeed the foolproof expression to be used for the winding number is $(\openone - h_V \tilde{g}^\dagger)$, whereas the $T$ matrix arises naturally from scattering theory and remains under normal conditions the object of choice.

\subsection{Case $y \neq 0$}

When $y\neq0$, the chiral decomposition of the Hamiltonian is\begin{gather}
    h^{top}(y)
	=-
	\left\{\openone+\bigl[\Tilde{g}(-y)\bigr]^{-1}\Tilde{T}^{-1\dagger}\bigl[\Tilde{g}(y)\bigr]^{-1}\Tilde{g}(0)\right\}^{-1}
\nonumber\\
	\times
    \bigl[\Tilde{g}(-y)\bigr]^{-1}\Tilde{T}^{-1\dagger}\bigl[\Tilde{g}(y)\bigr]^{-1},
\end{gather}
where $\Tilde{g}(y)$ is the chiral decomposition of $g(0,k_x,y)$. The winding number is then evaluated as before.
We obtain
\begin{gather}
    W(y)
	=
	\int_{-\pi}^{\pi}\frac{dk_x}{2\pi i}
\nonumber\\
	\times
	\partial_k
	\Bigl\{
		-\log\bigl[\det\bigl(
			\openone
			+
			\bigl[\Tilde{g}(-y)\bigr]^{-1}\Tilde{T}^{-1\dagger}\bigl[\Tilde{g}(y)\bigr]^{-1}\Tilde{g}(0)
		\bigr)
		\bigr]
\nonumber\\
		+
		\log\bigl[\det\bigl(\bigl[\Tilde{g}(-y)\bigr]^{-1}\bigr)\bigr]
		+
		\log\bigl[\det\bigl(\Tilde{T}^{-1\dagger}\bigr)\bigr]
\nonumber\\
		+
		\log\bigl[\det\bigl(\bigl[\Tilde{g}(y)\bigr]^{-1}\bigr)\bigr]
	\Bigr\},
\end{gather}
which gives Eq.~\eqref{eq:W(y)} in the main text.
If we consider the Lehmann representation of the Green's function \cite{economou_greens_2006}
\begin{gather}
    G(\omega,k_x)=\sum_n\frac{\ket{E_n(k_x)}\bra{E_n(k_x)}}{\omega_+-E_n(k_x)},
\end{gather}
where $E_n$ and $\ket{E_n}$ are the eigenvalues and eigenmodes of the Hamiltonian and $\omega_+=\omega+i0^+$, the local Green's function can be written as
\begin{gather}
	G(\omega,k_x,y,y)=\bra{y}G(\omega,k_x)\ket{y}
	\nonumber\\
	 =\sum_n\frac{\braket{y|E_n(k_x)}\braket{E_n(k_x)|y}}{\omega_+-E_n(k_x)}.
\end{gather}
For a trivial substrate the in-gap wavefunctions are localised to the interface region and normalisable. Therefore for such wavefunctions $\lim_{|y|\rightarrow\infty}\braket{y|E_{in-gap}(k_x)}=0$. Taking thus $|y|\to \infty$ for generic $V_m, k_m$ suppresses any pole from the in-gap modes. The remaining poles are from substrate and therefore gapped as the substrate is trivial.
This then means that $W(|y|\rightarrow \infty)=0$ and $H^{top}(k_x,|y|\rightarrow\infty)$ is trivial. Examining the form of the winding number we obtain
\begin{multline}
    W_{T^{-1}}
	=-
    \lim_{|y|\rightarrow\infty}\int_{-\pi}^{\pi}\frac{dk_x}{2\pi i}
\\
	\times\partial_k\log\Bigl\{
		\det\Bigl[
			\openone
			+
			\bigl[\Tilde{g}(-y)\big]^{-1}\Tilde{T}^{-1\dagger}\bigl[\Tilde{g}(y)\big]^{-1}\Tilde{g}(0)
		\Bigr]
	\Bigr\}.
\end{multline}
From the form of the Green's function in Eq.~(\ref{factorised-greens-function}) we see the term that gives the $y$ dependent winding number does not have any poles (as we have assumed the substrate is trivial). As such any change in the winding number as $y$ is varied must be due to zeros in the determinant of this term and so of $G(\omega=0,k_x,y,y)$.

\section{\label{appendix-proof-equivalence}Proof of the Equivalence of the Winding Numbers of $H$ and $H^{top}$}

In this section we prove that the winding numbers obtained for the full Hamiltonian $\mathcal{H}$
and for the topological Hamiltonian $H^{top}$ are equal, up to a fixed sign between them, and hence they
are equivalent descriptions of the topological properties of the system.

The substrate is assumed to be translationally invariant with a topologically trivial bulk Hamiltonian
$\mathcal{H}_0(k_x,k_y)$.
As the scattering interface breaks the translational symmetry in the $y$ direction we switch the description
from $k_y$ to positions $y$ and note that
the system then becomes one-dimensional with an effective unit cell that contains all $y$
labels. We shall denote the full Hamiltonian in this description $\mathcal{H}(k_x)$ and the
substrate Hamiltonian $\mathcal{H}_0(k_x)$. With the scattering included we have
$\mathcal{H}(k_x) = \mathcal{H}_0(k_x) + H_V(k_x) \otimes P_{y=0}$, where $P_{y=0}$ denotes the projection onto $y=0$.
The following proof will rely on the assumption stated in the main text that
$\det[g(\omega=0,k_x,y=0)] \neq 0$, and on the the already demonstrated
property that $W_{H^{top}}$ can only change when a pole of the $T$ matrix passes through $\omega=0$.
The topology will be characterised through the winding numbers for the different $k_x$ dependent
matrices from Hamiltonians or Green's functions.

The proof of the required theorem is broken up into several stages.

\begin{enumerate}[(a)]
\item
\textit{$\mathcal{H}_0(k_x)$ and $[g(\omega=0,k_x,y=0)]^{-1}$ are topologically trivial}
\end{enumerate}
We assume that the substrate is trivial, meaning that the relevant 2D topological index of $\mathcal{H}_0(k_x,k_y)$
is zero. This means there is a continuous deformation of $\mathcal{H}_0(k_x,k_y)$ to a trivial spectrally
flat ($k_x, k_y$ independent) Hamiltonian. Such a deformation preserves the gap at all $k_x,k_y$ and as such preserves
the gap of the 1D Hamiltonian $\mathcal{H}_0(k_x)$. This means that $\mathcal{H}_0(k_x)$ is continuously
connected to a trivial spectrally flat phase and so the winding number of $\mathcal{H}_0(k_x)$ is zero.

The eigenvalues of $\mathcal{H}_0(k_x)$ define the poles of $g(\omega=0,k_x,y=0)$ and so the zeros of
$[g(\omega=0,k_x,y=0)]^{-1}$. Since the considered deformation does not
produce a gap closure of $\mathcal{H}_0(k_x,k_y)$, it does not produce a gap closure of $[g(\omega=0,k_x,y=0)]^{-1}$ either.
Since we assume furthermore that initially $\det[g(\omega=0,k_x,y=0)] \neq 0$ it is possible to choose
a deformation that does not cause $\det[g(\omega=0,k_x,y=0)] = 0$ at any point, as this would require
for the assumed topologically trivial substrate a fine tuning of the parameters that is avoidable.
Hence we can choose a deformation that continuously connects $\mathcal{H}_0(k_x,k_y)$ to a $k_x$ independent
Hamiltonian while maintaining an invertible, finite $g(\omega=0,k_x,y=0)$ throughout. Hence
$[g(\omega=0,k_x,y=0]^{-1}$ transforms continuously to a $k_x$ independent matrix too,
showing that it is topologically trivial.

\begin{enumerate}[(b)]
\item
\textit{$\mathcal{H}(k_x)$ is trivial if and only if $H^{top}(k_x,y=0)$ is trivial}
\end{enumerate}

Let $\mathcal{H}(k_x)$ be topologically trivial. Then it can be continuously deformed to the $H_V = 0$
limit (i.e.\ to the substrate Hamiltonian), as we know it is topologically trivial too.
Such a deformation generates no gap closures in $\mathcal{H}(k_x)$ and so no poles of
$T(\omega = 0, k_x)$. Therefore the winding number of $H^{top}(k_x,y = 0)$ is preserved
and equal to winding number of $H^{top}(k_x,y = 0)|_{H_V=0}=-[g(\omega=0,k_x,y=0)]^{-1}$. From item (a) it follows that
$W_{H^{top}} = 0$.
Conversely, if $H^{top}(k_x,y=0)$ is topologically trivial it can be deformed continuously to the $H_V=0$ limit
without generation of any pole in $T(\omega=0,k_x)$.
Therefore $\mathcal{H}(k_x)$ can be continuously connected to $\mathcal{H}_0(k_x)$ which is trivial,
and consequently $W_{\mathcal{H}} = 0$.

\begin{enumerate}[(c)]
\item
\textit{$|W_{\mathcal{H}}| = |W_{H^{top}}|$}
\end{enumerate}

We shall show that the equality $|W_{\mathcal{H}}| = |W_{H^{top}}|$ follows from the simultaneous fulfilment
of the two inequalities
$|W_{\mathcal{H}}| \le |W_{H^{top}}|$ and $|W_{H^{top}}| \le |W_{\mathcal{H}}|$.
We make use of the fact that for any phase with a winding number of modulus $|W|$ deformations of the system
parameters to $W=0$ require at least $|W|$ gap closures, and that there are deformations that pass through
exactly $|W|$ gap closures. Let then $|W_\mathcal{H}| < |\mathcal{W}_{H^{top}}|$. Then there exists a deformation
of $\mathcal{H}(k_x)$ to the trivial $\mathcal{H}_0(k_x)$ by letting $H_V \to 0$ that generates only $|W_\mathcal{H}|$
gap closures, and as such exactly $|W_{\mathcal{H}}|$ poles of $T(\omega=0,k_x)$. However, since $|W_{H^{top}}|$ is
strictly larger than this number of gap closures, it is impossible that $H^{top}(k_x,y=0)$ could then connect
to the topologically trivial $H_V=0$ limit $H^{top}(k_x,y=0) = -[g(\omega=0,k_x,y=0)]^{-1}$. We must
conclude therefore that $|\mathcal{W}_{H^{top}}| \le |W_\mathcal{H}|$.
The converse is proven by swapping the roles of $\mathcal{H}(k_x)$ and $H^{top}(k_x,y=0)$.

\begin{enumerate}[(d)]
\item
\textit{$W_{\mathcal{H}} = \pm W_{H^{top}}$ with a fixed $\pm$ sign through the phase diagram}
\end{enumerate}
Consider a pair of Hamiltonians $\mathcal{H}_1(k_x)$ and $\mathcal{H}_2(k_x)$ together with their
associated topological Hamiltonians $H_1^{top}(k_x)$ and $H_2^{top}(k_x)$.
We assume first that $W_{\mathcal{H}_1} = - W_{\mathcal{H}_2} \neq 0$. We have already proven that then
$|W_{H_1^{top}}| = |W_{H_2^{top}}|$ but we wish to show furthermore that $W_{H_1^{top}} = - W_{H_2^{top}}$.
Let us assume that $W_{H_1^{top}} = W_{H_2^{top}}$ instead. Then there is a continuous deformation
between $H_1^{top}(k_x)$ and $H_2^{top}(k_x)$ that generates no gap closures and thus no poles of
$T(\omega=0,k_x)$. Consequently $\mathcal{H}_1(k_x)$ and $\mathcal{H}_2(k_x)$ must be continuously
connected too, which is a contradiction with the assumption $W_{\mathcal{H}_1} = - W_{\mathcal{H}_2}$.
Therefore we must have $W_{H_1^{top}} = - W_{H_2^{top}}$. The converse is proven by the same logic
as already applied above, by switching the role of $\mathcal{H}(k_x)$ and $H^{top}(k_x)$.
Note that we cannot say anything about the sign between $W_{\mathcal{H}}$ and $W_{H^{top}}$, but
as this sign is global to the entire phase diagram it is irrelevant.

As a consequence of items (a)--(d) the two winding numbers $W_{\mathcal{H}}$ and $W_{H^{top}}$
are equivalent, if not equal.

\section{Fourier transform of lattice s-wave superconductor Green's function}
\label{sec:Fourier}

The partially Fourier transformed retarded Green's function of a square lattice s-wave superconductor is evaluated. The Hamiltonian of such a 2D superconductor, after taking the gauge transformation $c^\dagger_{x,y,\sigma}\rightarrow e^{ik_m\sigma}c^{\dagger}_{x,y,\sigma}$, where $k_m$ is the spiral wavevector of the magnetic impurity interface, is given by
\begin{gather}
    \mathcal{H}_0(k_x,k_y)=\mathcal{H}^\uparrow\sigma^\uparrow+\mathcal{H}^\downarrow\sigma^\downarrow,
\end{gather}
where we define $\sigma^{\uparrow,\downarrow}=(\openone\pm\sigma^z)/2$, for $\sigma^z$ the Pauli-$z$ matrix acting on the spin degrees of freedom and
\begin{gather}
    \mathcal{H}^\sigma=\epsilon_{k_x,k_y,\sigma}\tau^z+\sigma\Delta\tau^x,
\end{gather}
where $\epsilon_{k_x,k_y,\sigma}=-2t(\cos(k_x+\sigma k_m)+\cos(k_y))-\mu$, the $\tau^i$ are the Pauli matrices acting on the particle-hole degrees of freedom, $t$ is the hopping integral, $\mu$ the chemical potential, and $\Delta$ the superconductor pairing amplitude chosen to be real and positive. The retarded Green's function of the superconductor is given by \cite{economou_greens_2006}
\begin{equation}
    g(\omega,k_x,k_y)=[\omega_+-\mathcal{H}_0(k_x,k_y)]^{-1},
\end{equation}where $\omega_+=\omega+i0^+$. The partial Fourier transform of the retarded Green's function is defined as \cite{carroll_subgap_2021}\begin{equation}
    g(\omega,k_x,y)=\int_{-\pi}^{\pi}\frac{dk_y}{2\pi}e^{ik_yy}g(\omega,k_x,k_y).
\end{equation}
For the given Hamiltonian we obtain
\begin{gather}
     \label{partially-fourier-greens-function-integral}g(\omega,k_x,y)=g^\uparrow(\omega,k_x,y)\sigma^\uparrow+g^\downarrow(\omega,k_x,y)\sigma^{\downarrow},
\end{gather}
with
\begin{gather}
    g^{\sigma}(\omega,k_x,y)
	=\frac{1}{2\pi}\int_{-\pi}^{\pi}dk_y\frac{e^{ik_yy}(\omega_+ + \epsilon_{k,\sigma}\tau^z+\sigma\Delta\tau^x)}{\omega_+^2-\epsilon_{k,\sigma}^2-\Delta^2}.
\end{gather}
We should notice that since $g(\omega,k_x,k_y)=g(\omega,k_x,-k_y)$ this integral is independent of the sign of $y$ and we can thus replace $y \to |y|$. We then change the integration variable to $z = e^{i k_y}$, with $dz=i z dk_y$, such that the integration is over the unit circle $S^1$ in the positive, anticlockwise direction, and obtain
\begin{equation}
    g^\sigma(\omega,k_x,y)=\frac{1}{2\pi i}\oint_{S^1}dz \, z^{|y|-1} \, g(\omega,k_x,z).
\end{equation}
As we focus on the in-gap states with $|\omega| < \Delta$, the small $+i0^+$ is of no importance we shall use $\omega$
instead of $\omega_+$ henceforth.
Focusing on just the denominator in the integrand
\begin{align}
	&z(\omega^2-\epsilon^{2}_{k_x,z,\sigma}-\Delta^2)
\nonumber\\
	&=
	z
	\Bigl\{\omega^2-\bigr[-t(z+1/z)-\mu_{k_x}\bigr]^2-\Delta^2\Bigr\}
\nonumber\\
	&=
	z (\omega^2-\Delta^2)-\frac{z}{z^2}\bigl[t(z^2+1)+\mu_{k_x}z\bigr]^2
\nonumber\\
    &=
	\frac{t^2}{z}
	\Bigl\{ z^2(\Tilde{\omega}^2-\Tilde{\Delta}^2)-\bigl[(z^2+1)+\Tilde{\mu}_{k_x}z\bigr]^2\Bigr\},
\end{align}
where we have defined $\Tilde{\mu}_{k_x}=\Tilde{\mu}+2\cos(k_x+\sigma k_m)$ and introduced the dimensionless
energies $\tilde{\omega}=\omega/t$, $\tilde{\Delta}=\Delta/t$, and $\tilde{\mu} = \mu/t$.
The integral then becomes
\begin{align}
	&g^\sigma(\omega,k_x,y)
	=
\nonumber\\
	&\frac{1}{2\pi it^2}\oint_{S^1}
	\frac{dzz^{|y|+1}g^\sigma(\omega,k_x,z)}{z^2(\Tilde{\omega}^2-\Tilde{\Delta}^2)-[(z^2+1)+\Tilde{\mu}_{k_x}z]^2}.
\end{align}
The numerator is given by
\begin{multline}
    z^{|y|+1}g^\sigma(\omega,k_x,z)=\\
	tz^{|y|}\left[\Tilde{\omega}z-(\Tilde{\mu}_{k_x}z+z^2+1)\tau^z+z\sigma\Tilde{\Delta}\tau^x\right].
\end{multline}
All the poles are contained within the denominator and we can safely proceed by solving
\begin{gather}
    0=z^2(\Tilde{\omega}^2-\Tilde{\Delta}^2)-\bigl[(z^2+1)+\Tilde{\mu}_{k_x}z\bigr]^2,
\end{gather}
or
\begin{gather}
    z^2(\Tilde{\omega}^2-\Tilde{\Delta}^2)=\bigl[(z^2+1)+\Tilde{\mu}_{k_x}z\bigr]^2,
\end{gather}
from which the square root leads to the two quadratic equations
\begin{gather}
    0=z^2+z(\Tilde{\mu}_{k_x}\pm\sqrt{\Tilde{\omega}^2-\Tilde{\Delta}^2})+1.
     \label{greens-function-pole}
\end{gather}
We thus obtain the four roots of the denominator
\begin{multline}
    z_{\pm_1,\pm_2}
    =\frac{1}{2} \Bigl[\Tilde{\mu}_{k_x}\pm_1\sqrt{\Tilde{\omega}^2-\Tilde{\Delta}^2}\Bigr]
\\
	\pm_2 \frac{1}{2} \sqrt{\left(-\Tilde{\mu}_{k_x}\pm_1\sqrt{\Tilde{\omega}^2-\Tilde{\Delta}^2}\right)^2-4}.
\end{multline}
Then the integral is
\begin{gather}
    g^{\sigma}(\omega,k_x,y)=
\nonumber\\
	\frac{-1}{2\pi i t}\oint_{S^1}\frac{z^{|y|}dz}{(z-z_{+,+})(z-z_{-,+})(z-z_{+,-})(z-z_{-,-})}
\nonumber\\
	\times\left[\Tilde{\omega}z-(\Tilde{\mu}_{k_x}z+z^2+1)\tau^z+z\sigma\Tilde{\Delta}\tau^x\right].
\end{gather}
The total minus sign arises from the negative $z^4$ coefficient in the denominator. The integral can then be evaluated using Cauchy's residue theorem.
\begin{figure}
    \centering
    \includegraphics[width=0.5\textwidth]{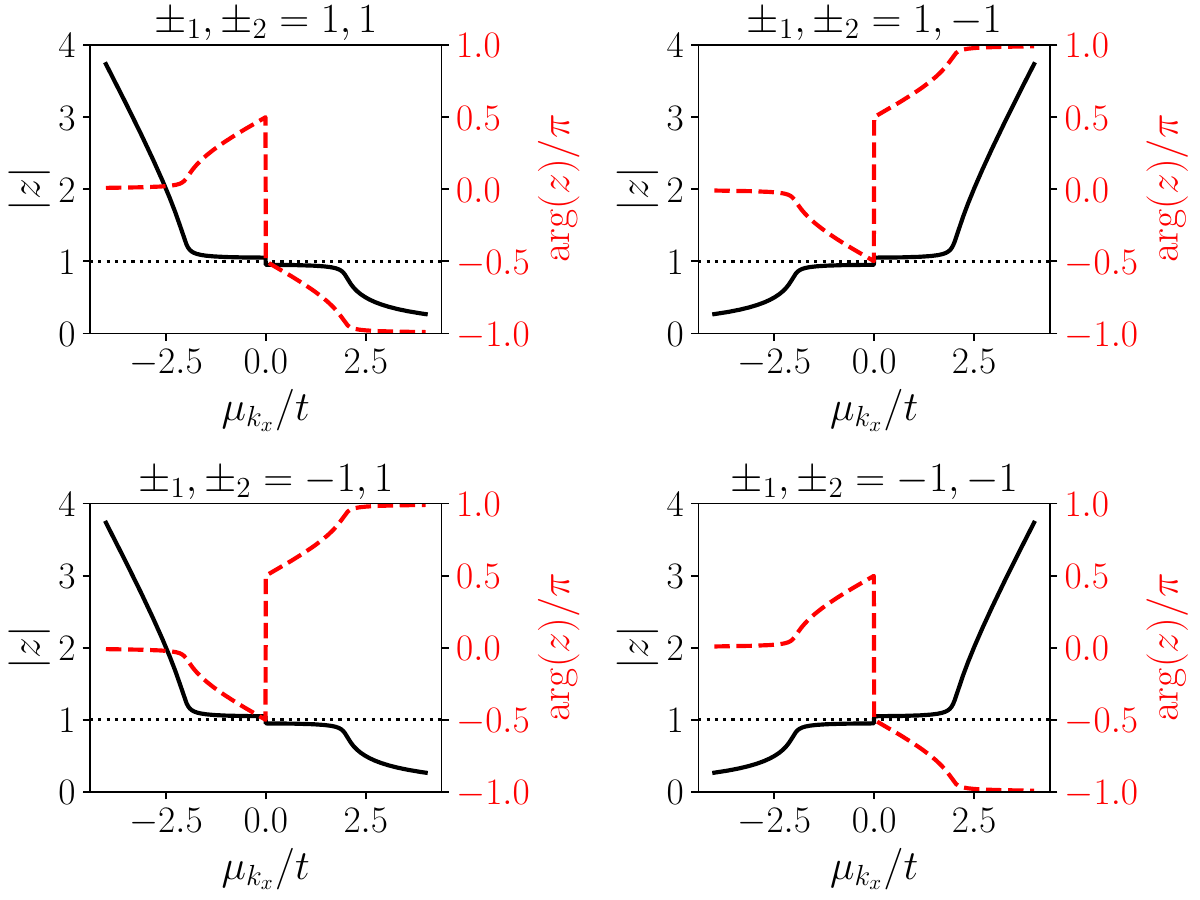}
    \caption{Modulus (black, solid) and phase (red, dashed) of the poles obtained from Eq.~\eqref{greens-function-pole} for $\omega=0$ and $\Delta=0.1t$. The black dotted line shows $|z|=1$. One can verify that as long as $|\omega|<\Delta$, the two poles $z_{\pm,\sign(\mu_{k_x})}$ are within in the unit circle.}
    \label{Fig:pole-location}
\end{figure}
In Fig.~\ref{Fig:pole-location} we show the value of the poles for the example of $\omega=0$ as a function of $\mu_{k_x}$.
For $|\omega| < \Delta$ the poles $z_{1,2} = z_{\pm,\sign(\mu_k)}$ lie within the unit circle, and the poles $z_{3,4} = z_{\pm,-\sign(\mu_k)}$ are outside the unit circle.
The Green's function then evaluates to
\begin{gather}
    \label{sc-greens-function}
    g^\sigma(\omega,k_x,y)=\frac{-\xi_-(\Tilde{\omega}+\sigma\Tilde{\Delta}\tau^x)+i\xi_+\sqrt{\Tilde{\Delta}^2-\Tilde{\omega}^2}\tau^z}{t(z_1-z_2)}
\end{gather}
where
\begin{gather}
    \xi_{\pm}=\frac{z_1^{|y|+1}}{(z_1-z_3)(z_1-z_4)}\pm \frac{z_2^{|y|+1}}{(z_2-z_3)(z_2-z_4)}.
\end{gather}
We note that $\xi_+$ is real and $\xi_-$ purely imaginary as $z_1=z_2^*$ and $z_3=z_4^*$. Consequently $g^\sigma$ is real as expected for a Green's function when $\omega$ lies in a range with vanishing density of states.
The $\xi_\pm$ functions are non-singular as long as $\Delta\neq0$. We emphasise that this is the exact Green's function of a square lattice tight binding s-wave superconductor, and no low-energy approximation has been made.

\begin{figure*}[t]
    \centering
    \includegraphics[width=0.8\textwidth]{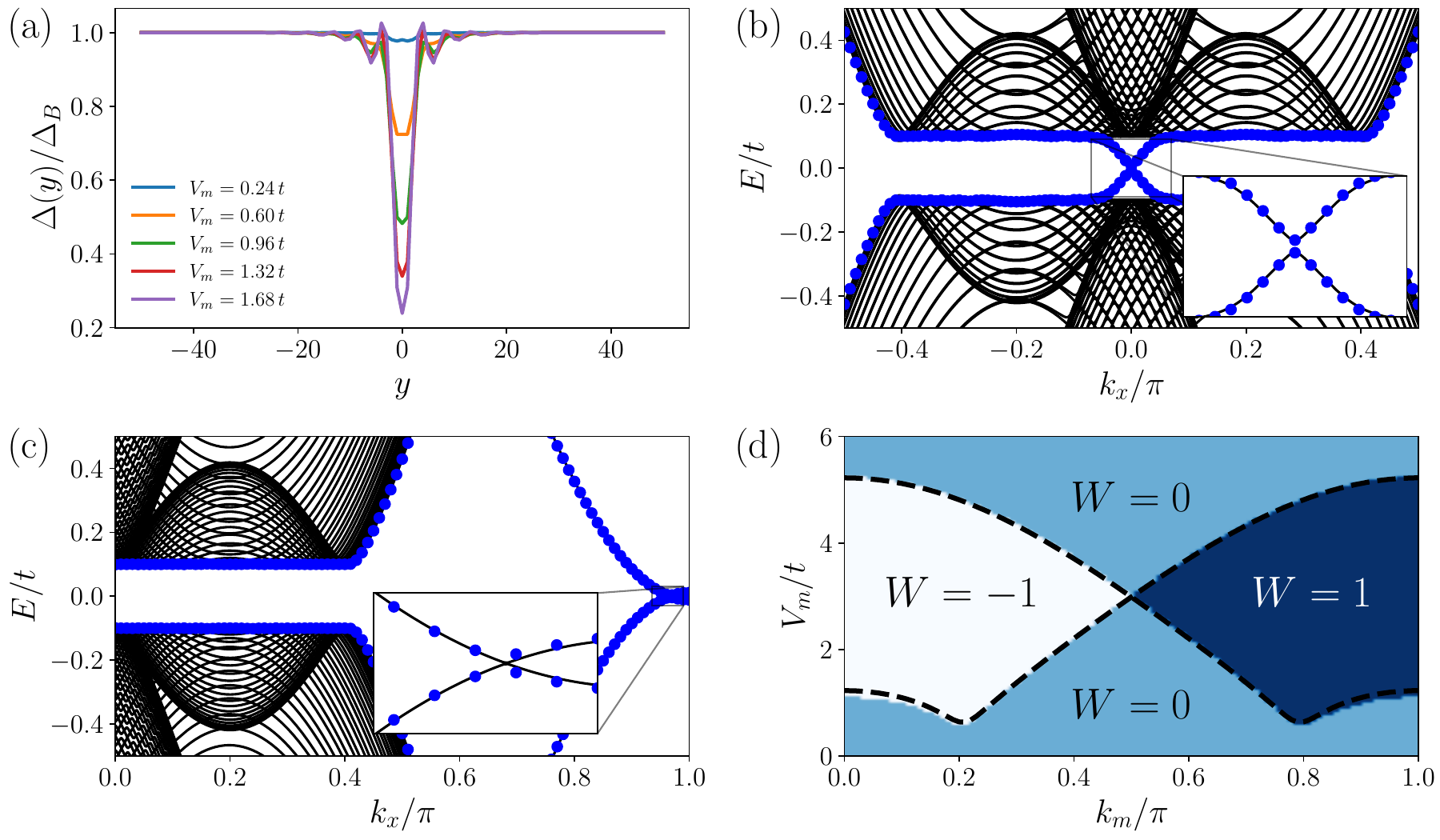}
    \caption{Results of self-consistent calculations for $N_y=101$, $\Delta_B=0.1t$ and $k_F=0.65$. Panel (a) Shows $\Delta(y)$ for $k_m=k_F$ and various values of $V_m$. Although the scattering potential has a considerable effect at the impurity site at $y=N_y/2$, the renormalisation of $\Delta(y)$ almost immediately drops to below 10\% independently of the scattering strength. Panels (b) and (c) show a comparison of the band structure of the self-consistent solution (black lines) and lowest two eigenvalues of the non-self consistent $\Delta(y)=\Delta_B$ (blue circles) for $k_m=0.63$, and $V_m=0.6$ and $4.8$ respectively. There is very little difference between the two, particularly around the gap closures at $k_x=0,\pi$. The insets confirm that there is little change in the electronic structure. Panel (d) shows the phase diagram obtained from Eq.~\eqref{eq:nu} together with the analytically calculated phase boundaries from Eq.~(\ref{phase-boundaries}). There is almost perfect agreement. }
    \label{Fig:self-consistent-restults}
\end{figure*}

\section{Phase Diagram}
\label{sec:phase_diagram}

With the explicit form of the Green's function for the impurity chain given by Eq.~\eqref{Greens-function-dyson-series} in the main text, we can solve for the conditions of a phase boundary by finding poles of the $T$ matrix at $\omega=0$,
\begin{equation}
    \det\bigl[T^{-1}(0,k_x)\bigr]=0.
    \label{phase-boundary-definition}
\end{equation}
This equation can be simplified by making use the model's chiral symmetry at $\omega=0$. We can write the $T$ matrix in the eigenbasis of the chiral symmetry, using the unitary
\begin{equation}
    U=\frac{1}{\sqrt{2}}\begin{pmatrix}
        1&0&1&0\\
        0&i&0&i\\
        0&-1&0&1\\
        i&0&-i&0\\
    \end{pmatrix},
\end{equation}
written in the basis
\begin{equation}
	(c_{k_x,y=0,\uparrow},c_{k_x,y=0,\downarrow},c^\dagger_{-k_x,y=0,\uparrow},c^\dagger_{-k_x,y=0,\downarrow}).
\end{equation}
This transforms the $T$ matrix to
\begin{equation}
    U^\dagger T^{-1}U=\begin{pmatrix}
        0&\Tilde{T}^{-1}\\
        \Tilde{T}^{-1\dagger}&0
    \end{pmatrix},
\end{equation}
with the notation $T=T(\omega=0,k_x)$ and \begin{equation}
    \det(T^{-1})=|\det(\Tilde{T}^{-1})|^2.
\end{equation}
Defining the notation $g_{ij}^\sigma=[g^\sigma(\omega=0,k_x,y=0)]_{ij}$ and using that at $\omega=0$ we have $g^\sigma_{11}=-g^\sigma_{22}$ and $g^\sigma_{12}=g^\sigma_{21}$, we obtain
\begin{equation}
    \Tilde{T}^{-1}=\begin{pmatrix}
        ig^\uparrow_{12}-g^\uparrow_{11}&\frac{i}{V_m}\\
        -\frac{i}{V_m}&ig^\downarrow_{12}-g_{11}^\downarrow
    \end{pmatrix}.
    \label{T-matrix-chiral-decomposition}
\end{equation}
The condition for a phase transition is $\det(T^{-1})=0$, which becomes
\begin{gather}
	0=g^\uparrow_{11}g^\downarrow_{11}-g_{12}^{\uparrow}g_{12}^\downarrow-i(g_{11}^\uparrow g_{12}^\downarrow+g^\downarrow_{11}g^\uparrow_{12})-\frac{1}{V_m^2}.
\end{gather}
The bare Green's function is real as the density of states is zero at $\omega=0$, and as noted earlier we can omit the infinitesimal shifts $i0^+$. Then the real and imaginary parts split into two equations. The imaginary part gives
\begin{equation}
    g^\downarrow_{11}g^\uparrow_{12}=-g^\uparrow_{11}g^\downarrow_{12}.
\end{equation}
If $k_m\neq n\pi$ for $n\in\mathbb{Z}$, this has solutions for any set of system parameters, but required $k_x=0,\pi$. This is because the only differences between the two spin species is the kinetic energy $\cos(k_x+\sigma k_m)$ and the sign of the pairing $\sigma\Delta$. At $k_x=0,\pi$, the dispersions are equal as $\cos(k_x+\sigma k_m)=\pm\cos(\sigma k_m)=\pm\cos(k_m)$. Therefore $g^\uparrow_{11}=g^\downarrow_{11}$ and $g^\uparrow_{12}=-g^\downarrow_{12}$, and so the above equation is satisfied. When $k_m=0,\pi$, the above equation is satisfied for any $k_x$ meaning the gap closes and stays closed rather than opening up again.

The real part gives the conditions for the phase transition and along with the symmetry of the spin species at $k_x=0,\pi$. We find
\begin{gather}
    V^*_{m,k_x}=\pm\left\{[g^{\uparrow}_{11}(k_x)]^2+[g_{12}^{\uparrow}(k_x)]^2\right\}^{-1/2}.
    \label{app:phase-boundaries}
\end{gather}
giving Eq.~\eqref{phase-boundaries} in the main text. As gap closures only occur at $k_x=0,\pi$, the winding number is restricted to $W=0,1,-1$ and therefore it is sufficient to find its parity. This can written as the sign of the product of the determinants of the chiral decomposition at $k_x=0,\pi$ \cite{tewari_topological_2012}. Using the chiral decomposition above, this is given by
\begin{gather}
    (-1)^{W(y=0)}=\prod_{k_x=0,\pi}\sign\left[({g^{\uparrow}_{11}})^2+({g^{\uparrow}_{12}})^2-({V}_m)^{-2}\right]
\nonumber\\
   =\prod_{k_x=0,\pi}\sign\left[(V^*_{m,k_x})^{-2}-(V_m)^{-2}\right],
\end{gather}
giving Eq.~\eqref{eq:parity} in the main text.

\section{Phase Diagram of Self-Consistent Solution}
\label{sec:self_consistent}

We describe a self-consistent solution to the superconducting gap equations in the presence of a magnetic interface and show that it makes very little difference to the overall phase diagram. The self-consistency condition is the same that was used in Ref.~\cite{carroll_subgap_2021,carroll_subgap_2021-1}. We restrict ourselves to just s-wave pairing as the triplet pairing is extremely weak. We maintain translational invariance along the $x$-direction. Then the self consistency condition is
\begin{gather}
    \Delta(y)=V_P\braket{c^\dagger_{x,y,\uparrow}c^\dagger_{x,y,\downarrow}}
	=V_P\sum_{k_x}\braket{c^\dagger_{k_x,y,\uparrow}c^\dagger_{-k_x,y,\downarrow}},
\end{gather}
where $\braket{\dots}$ denotes the ground state average.
The translational invariance along the $x$-direction means the pairing is only a function of $y$. The pairing potential strength $V_P$ is chosen such that the pairing tends to the chosen bulk pairing, $\Delta_B$ (i.e. the pairing at $V_m=0$)
\begin{equation} \label{eq:VP}
	V_P=\frac{\Delta_B}{\sum_{k_x}\braket{c^\dagger_{k_x,|y|\rightarrow\infty,\uparrow}c^\dagger_{-k_x,|y|\rightarrow\infty,\downarrow}}}.\end{equation}
Starting from the initial guess $\Delta(y)=\Delta_B$ we calculate $\braket{c^\dagger_{k_x,y,\uparrow}c^\dagger_{-k_x,y,\downarrow}}$ from direct diagonalisation of the tight-binding matrix. With this result we update $V_P$ through Eq.~\eqref{eq:VP} to guarantee that $\Delta(y)$ tends to $\Delta_B$ at large $y$. With this we reconstruct the Hamiltonian and iterate until convergence. With the final solution of the tight-binding model we calculate a topological invariant of the bands. As it is computationally heavy to calculate the winding number of a large matrix, we make use of the inversion symmetry to calculate an equivalent invariant. This is defined as
\begin{equation} \label{eq:nu}
    \nu=n_0-n_\pi,
\end{equation}
where $n_{k_x}$ is the number of negative non-zero eigenvalues of $P(k_x)\mathcal{I}P(k_x)$ where $\mathcal{I}$ is the inversion symmetry operator. This has been shown to be equal to the number of $\zeta=1/2$ modes in the entanglement spectrum localised to the edge of the system \cite{hughes_inversion-symmetric_2011}.

The self consistent solution to the gap is shown in Fig.~\ref{Fig:self-consistent-restults}(a) for various $V_m$. We see that the gap is lowered by around 20\% around the magnetic impurity around the critical scattering $V_{m,k_x=0}^*\approx0.63t$. The residual error of this solution is on the order of $10^{-8}\Delta_B$. We show the band structure for two values of $V_m$ around the phase boundaries in Fig.~\ref{Fig:self-consistent-restults}(b)--(c) and compare it to the in-gap bands of the non-self-consistent solution. We again see that around the bottom of the band very little difference is made and the only visible deviations are around the $k_x\approx 2k_F$ where the in-gap bands of the self-consistent solution are slightly lower than the non-self-consistent one. Finally, in Fig.~\ref{Fig:self-consistent-restults}(d) we show the phase diagram of the invariant $\nu$ defined above with the non-self-consistent phase boundaries overlayed and again we see perfectly agreement up to the pixel size.
We conclude that self-consistency has so little effect on all relevant results such that it can be safely ignored.



%

\end{document}